\documentclass[11pt]{amsart}
\usepackage{graphicx}
 \usepackage{geometry}  
\usepackage{palatino, url, multicol}
\usepackage{amsmath, amsthm}
\usepackage{fullpage}
\usepackage{pdflscape}
\usepackage{fancyhdr}
\usepackage{verbatim} 
\usepackage{wrapfig}
\usepackage{booktabs}
\DeclareMathAlphabet{\mathpzc}{OT1}{pzc}{m}{it}
\title{Experimental and Semi-Empirical Branching Fractions\\ of the $ \bf3\text{s}^23\text{p}^2-3\text{s}3\text{p}^3$ J=2 Transition Array in P II}
\author{Jeremy Bancroft Brown$^a$, Lorenzo J. Curtis$^b$, and David G. Ellis$^b$\\\tiny $^a$Department of Physics, University of Chicago, Chicago IL 60637\\$^b$Department of Physics and Astronomy, University of Toledo, Toledo OH 43606\\NSF Physics REU Program at the University of Toledo, Toledo OH 43606}
\date{August 30, 2008}
\begin{document}
\begin{abstract}
A semi-empirical method is used to characterize the $ 3\text{s}^23\text{p}^2-3\text{s}3\text{p}^3 $ J=2 transition array in P II. In this method, Slater, spin-orbit, and radial parameters are fitted to experimental energy levels in order to obtain a description of the array in terms of LS-coupling basis vectors. The various IC and CI amplitudes resulting from this model are then used to predict the branching fractions of transitions within the array. Results close to LS-coupling values are presented, and these are compared to branching ratios measured using beam-foil spectroscopy at the THIA laboratory. The work provides support for the hypothesis of Dr. Curtis that transition arrays with little upper state IC but significant upper state CI in atoms of low Z exhibit branching fractions close to LS-coupled values, although the data are inconclusive in this respect.
\end{abstract}
\maketitle
\fancypagestyle{plain}{
\pagestyle{fancy}
\lhead{} 
\chead{} 
\rhead{} 
\lfoot{} 
\cfoot{} 
\fancyfoot[LE,RO]{\thepage}
\renewcommand{\headrulewidth}{0.0pt} 
\renewcommand{\footrulewidth}{0.0pt} 
}
\thispagestyle{plain}
\pagestyle{plain}
\begin{multicols}{2}
\section{Introduction}
In light atoms carrying a small central charge, relativistic effects upon the electron cloud are generally small. For this reason, it is convenient to refer to levels within such systems using the LS coupling model. However, in any multi-electron atomic system, LS coupling fails as a precise model due to direct and exchange coulomb interactions between various electrons, relativistic spin-orbit effects, and, somewhat less significantly, the magnetic dipole-dipole interaction between each electron and the nucleus. In order to take these interactions into account, the intermediate coupling (IC) model must be used. In this approximation, each electron is described by a distinct wavefunction that is a linear combination of LS basis states with different $L$ or $S$, but the same $J$. The IC model preserves the convenience of utilizing LS selection rules when identifying transitions, and it can be generalized to include configuration interaction (CI), in which the orthonormal basis set is extended to comprise LS states from different configurations. Because of this, the IC/CI model provides a useful picture of the atom that can be used to make predictions and isolate trends along isoelectronic sequences.

In order to successfully use the IC/CI model to characterize an atomic system, it is necessary to obtain the mixing amplitudes that give the electronic eigenstates of the atom in terms of an LS basis set. These amplitudes may be determined \emph{ab initio}, using iterative energy-minimization algorithms such as Multi-Configuration Hartree Fock (MCHF), or they can be arrived at semi-empirically \cite{cowan1981}. In this paper, we use the latter approach, which relies upon the high accuracy and precision of NIST energy-level data to obtain accurate results. \section{The Semi-Empirical Method}In the semi-empirical technique we use to model the system, the energies contained within the various interactions described above are treated as parameters and fitted to measured energy-level data. These parameters include the Slater direct and exchange energies $F^k$ and $G^k$, the spin-orbit energies $\zeta_k$, and the radial configuration interaction energy $R_k$. Linear combinations of the parameters are placed in a symmetric nondiagonal interaction matrix expressed in the finite LS basis of interest, where each matrix element corresponds to the Hamiltonian between two LS states. IC and CI mixing amplitudes are obtained by diagonalizing the matrix and finding its eigenvectors, which correspond to single-electron wavefunctions in terms of LS basis states. 

In order to diagonalize the matrix, we obtain a characteristic equation of the form 
\begin{equation*} \displaystyle P_N(\lambda) = \sum_{n=0}^N a_n\lambda^n = 0 \end{equation*}
where each $a_n$ is expressed in terms of the various energy parameters. The roots of this equation are the eigenvalues of the interaction matrix, which would ideally correspond closely to the measured energy levels of the system. However, due to the imperfect nature of the model, it is generally impossible to find real parameter values which give this result. Instead, we use the method of Z. B. Rudzikas \cite{rudzikas1991}, noting that because the characteristic equation can be factored into its experimentally measured roots,
\begin{equation*} \displaystyle (\lambda-E_1)(\lambda-E_2)(...)(\lambda-E_N)= 0 \end{equation*}
the coefficients of $\lambda$ can be expressed as sums and products of experimental energies. For instance: \cite{keller1946}
\begin{equation*} \begin{gathered} \displaystyle a_{N-1}=-\sum_{i=1}^N E_i \\
a_{N-2}=\sum_{i=1}^N\sum_{j>i}^N E_iE_j \\
\vdots \\
a_{0}=(-1)^N\prod_{i=1}^NE_i\end{gathered} \end{equation*} \\
\thispagestyle{empty} 
Thus, the energy parameters may be adjusted so that the $a_n$ in the characteristic equation fit these experimental coefficients. This is accomplished using a numerical least-squares algorithm that searches parameter space for the global minimum of a $\chi^2$ surface. When such a minimum is encountered, its parameter coordinates are printed and the interaction matrix may be expressed numerically. The eigenvalues and eigenvectors of the interaction matrix can then be found, giving the IC and CI mixing amplitudes that facilitate semi-empirical computation of branching fractions and lifetimes.\\ \section{Application of the Semi-Emprical Method to J=2 Transitions in P II} Here, we have applied this method to  $ 3\text{s}^23\text{p}^2-3\text{s}3\text{p}^3$ J=2 transitions in P II. Due to significant configuration interaction between the $3\text{s}3\text{p}^3$ and $3\text{s}^23\text{p}3\text{d}$ upper states, it was necessary to devise an 8x8 interaction matrix for J=2 that would take the levels of both of these upper configurations into account (see Figures 1-3). This matrix included the average configuration energies $E_A$ and $E_B$ in addition to the other parameters mentioned earlier, making ten parameters total. The coefficients of the various parameters were determined from tables in Condon and Shortley \cite{condonshortley1935}. The model was underdetermined, so in order make the system overdetermined and "steer" the $\chi^2$ fit into a physical result with some empirical and \emph{ab initio} information, $E_A$ and $E_B$ were set to the measured average configuration energies, while $R_1$, $\zeta_p$, and $\zeta_d$ were fixed to MCHF\footnote{Note that the p shell is half-filled in the  $3\text{s}3\text{p}^3$ configuration, so $\langle \hat{H}_\text{so} \rangle = 0$,  causing $\zeta_{\text{p}}$ to appear only as a second-order, off-diagonal parameter in the interaction matrix.} values: \\ 
\begin{center}
\begin{tabular}{c r}
\hline\hline
Parameter&Value (cm$^{-1}$)\\ [0.0ex]
\hline
$E_A$ & $78,700$  \\
$E_B$ & $96,500$  \\
$R_1$&$13,348$ \\
$\zeta_p$&$295$ \\
$\zeta_d$&$0$ \\ \hline\hline \\
\end{tabular}
\end{center}

Due to fixing these five parameters, the actual parameter space was $\mathbb{R}^5$. The $\chi^2$ fit was accomplished using Mathematica; however, the software's nonlinear regression package would only fit a single analytic function to a set of data points, while in our case it was necessary to fit eight distinct expressions simultaneously. For this reason, it was necessary to write out a manual weighted computation of $\chi^2$ and then use the internal function NMinimize to search the parameter space (see Figure 4).
\end{multicols}
\begin{figure}[h]

\begin{small} $\ \left \{ \begin{array}{llll}\text{1: }\mid 3\text{s}3\text{p}^3\text{ }{}^5\text{S}_2^{\text{o}} \rangle & \text{2: } \mid 3\text{s}3\text{p}^3\text{ }{}^3\text{D}_2^{\text{o}} \rangle & \text{3: } \mid 3\text{s}3\text{p}^3\text{ }{}^3\text{P}_2^{\text{o}} \rangle & \text{4: } \mid 3\text{s}3\text{p}^3\text{ }{}^1\text{D}_2^{\text{o}} \rangle \\ \text{5: } \mid 3\text{s}^{2}3\text{p}3\text{d}\text{ }{}^1\text{D}_2^{\text{o}} \rangle & \text{6: } \mid 3\text{s}^{2}3\text{p}3\text{d}\text{ }{}^3\text{F}_2^{\text{o}} \rangle & \text{7: } \mid 3\text{s}^{2}3\text{p}3\text{d}\text{ }{}^3\text{P}_2^{\text{o}} \rangle & \text{8: } \mid 3\text{s}^{2}3\text{p}3\text{d}\text{ }{}^3\text{D}_2^{\text{o}} \rangle \end{array} \right \} $\ \end{small}
\caption{The ordered LS basis set of the J=2 interaction matrix}
\end{figure}
\begin{figure}[h]
\centering
\begin{small}
\begin{center}
$\ \left(\begin{array}{cccccccc}
  \Lambda_1& 0 & \zeta_p & 0 & 0 & 0 & 0 & 0 \\
 0 & \Lambda_2& \frac{\sqrt{3} \zeta_p}{2} & 0 & 0 & 0 & 0 & \frac{R_1}{\sqrt{5}} \\
 \zeta_p & \frac{\sqrt{3} \zeta_p}{2} & \Lambda_3& \frac{\zeta_p}{\sqrt{2}} & 0 & 0 & -\frac{R_1}{3} & 0 \\
 0 & 0 & \frac{\zeta_p}{\sqrt{2}} &  \Lambda_4& -\frac{R_1}{\sqrt{5}}& 0 & 0 & 0 \\
 0 & 0 & 0 & -\frac{R_1}{\sqrt{5}} & \Lambda_5& -\sqrt{\frac{7}{30}} (\zeta_d+\zeta_p) & \frac{3 (\zeta_d+\zeta_p)}{2 \sqrt{10}} & \frac{-5 \zeta_d+\zeta_p}{2 \sqrt{6}} \\
 0 & 0 & 0 & 0 & -\sqrt{\frac{7}{30}} (\zeta_d+\zeta_p) & \Lambda_6& 0 & -\frac{1}{3} \sqrt{\frac{7}{5}} (-\zeta_d+\zeta_p) \\
 0 & 0 & -\frac{R_1}{3} & 0 & \frac{3 (\zeta_d+\zeta_p)}{2 \sqrt{10}} & 0 & \Lambda_7& -\frac{3}{4} \sqrt{\frac{3}{5}} (-\zeta_d+\zeta_p) \\
 0 & \frac{R_1}{\sqrt{5}} & 0 & 0 & \frac{-5 \zeta_d+\zeta_p}{2 \sqrt{6}} & -\frac{1}{3} \sqrt{\frac{7}{5}} (-\zeta_d+\zeta_p) & -\frac{3}{4} \sqrt{\frac{3}{5}} (-\zeta_d+\zeta_p) & \Lambda_8
\end{array}\right) $\ \\
\end{center}
\end{small}
\caption{The J=2 interaction matrix}
\end{figure}
\begin{figure}[h]
\begin{small}
\begin{flalign*} \Lambda_1 & = E_A-\frac{9 F^2(3\text{p},3\text{p})}{25}-\frac{G^1(3\text{s},3\text{p})}{2} \\ \Lambda_2 & = E_A-\frac{G^1(3\text{s},3\text{p})}{6}\\ \Lambda_3 & = E_A+\frac{6 F^2(3\text{p},3\text{p})}{25}-\frac{G^1(3\text{s},3\text{p})}{6} \\ \Lambda_4 & = E_A+\frac{G^1(3\text{s},3\text{p})}{2} \\ \Lambda_5 & = E_B-\frac{F^2(3\text{p},3\text{d})}{5}-\frac{2 G^1(3\text{p},3\text{d})}{15}+\frac{9 G^3(3\text{p},3\text{d})}{70}\\ \Lambda_6 & = E_B+\frac{2 F^2(3\text{p},3\text{d})}{35}-\frac{G^1(3\text{p},3\text{d})}{3}+\frac{3 G^3(3\text{p},3\text{d})}{98}-\frac{2}{3} (2 \zeta_d+\zeta_p) \\ \Lambda_7 & = E_B+\frac{F^2(3\text{p},3\text{d})}{5}-\frac{3 G^3(3\text{p},3\text{d})}{14}+\frac{1}{4} (3 \zeta_d-\zeta_p)\\ \Lambda_8 & = E_B-\frac{F^2(3\text{p},3\text{d})}{5}+\frac{4 G^1(3\text{p},3\text{d})}{15}-\frac{3 G^3(3\text{p},3\text{d})}{70}+\frac{1}{12} (-5 \zeta_d-\zeta_p) \end{flalign*} \end{small}
\caption{Diagonal elements of the J=2 interaction matrix}
\end{figure}
\begin{figure}[h]
\begin{verbatim}
chiSquared = 0;
For[i = 1, i < 9, i++, chiSquared = chiSquared + ((energyCoefficients[[i]] 
- Part[lambdaCoefficients, 9 - i])^2)/(energyCoefficients[[i]]^2)];
\end{verbatim}
\caption{Manual computation of $\chi^2$ in Mathematica\setcounter{footnote}{\value{footnote}}\protect\footnotemark
}
\end{figure}
\footnotetext{To weight $\chi^2$ for uncertainty, the code divides each squared difference by the square of the corresponding numerical coefficient of $\lambda$. This is based upon the assumption that the uncertainties in the measured energies are essentially proportional to their magnitude. }
\begin{multicols}{2}
The parameter search ultimately returned $\chi^2 \approx 2.76893\times10^{-13}$ and gave the following values for each of the five Slater parameters:\\
\begin{center}
\begin{tabular}{c r r}
\hline\hline
Parameter&$\chi^2$ fit (cm$^{-1}$)&MCHF (cm$^{-1}$)\\ [0.0ex]
\hline
$F^2$(3p, 3p) & $9,646$  &$48,076$\\ 
$F^2$(3p, 3d) & $48,202$  &$5,334$\\
$G^1$(3p, 3d) &  $6,643$ &$2,824$\\
$G^1$(3s, 3p) &$59,117$ &$66,845$\\
$G^3$(3p, 3d) &$-22,040$ &$1,984$\\ \hline\hline \\
\end{tabular}
\end{center}
While both of the $G^1$ parameters exhibit order-of-magnitude agreement with MCHF values, both of the $F^2$ parameters and the $G^3$ parameter differ significantly from the MCHF calculation. However, this comparison may not be as unfavorable as it first appears. It is worth noting that at the MCHF parameter coordinates, the height of the $\chi^2$ surface is $0.118162$. Thus, while this value is many orders of magnitude larger than the best-fit value of $2.76893\times10^{-13}$, the 8x8 interaction matrix model is a good fit to measured energy-level data even with MCHF parameters. This is an encouraging conclusion, because MCHF is considerably more sophisticated than a nonlinear regression technique in this context and has been more reliably applied to complex atomic systems \cite{fischer2006}. Therefore, although the 8x8 matrix model returns results considerably different from MCHF, its characterization of the system is approximately physical. 

It is difficult to construct a conceptual link between the mechanisms of a parameter-search procedure and the physical reality of an atomic system; however, some tentative comments may be made on these results.  First of all, it is not certain that the disagreement between the MCHF results and the $\chi^2$ results is entirely due to imperfection in the 8x8 model; it is possible for the 8x8 model to capture effects that MCHF does not. The apparent "switching" of the $F^2$ parameters is of particular interest in this respect. Basic physical intuition supports the MCHF result that $F^2$(3p, 3p) $>$ $F^2$(3p, 3d), because it is reasonable to expect that $\langle R_{3d} \rangle > \langle R_{3p} \rangle$, reducing the p/d coulomb interaction relative to the p/p interaction. Even so, the MCHF algorithm utilizes a central-field approximation that can leave out the electron correlation effects inherent in the actual multi-electron wavefunction of the system. Thus, the $\chi^2$ characterization of these parameters could indicate the presence of electron correlation and markedly distinct angular expectation values among the electrons of the 3p suborbital\footnote{This was suggested by Dr. Ellis.}. However, the results in this regard are highly inconclusive and such an interpretation represents little more than conjecture at this time.

A comparison of fitted and experimental energy levels provides another test of the $\chi^2$ fit quality. In order to produce this comparison, eigenvalues and their corresponding eigenvectors were computed from the numerical J=2 matrix. LS states and eigenvectors were then assigned to each other in pairs according to the largest component of each eigenvector.
\begin{center}
\begin{tabular}{r@{      }l r r}
\hline\hline
Config. &Level&$E_{\text{Fitted}}$(cm$^{-1}$)&$E_{\text{Exp.}}$ (cm$^{-1}$) \cite{nist}\\ [0.0ex]
\hline
$3\text{s}3\text{p}^3$&$ {}^5\text{S}_2^{\text{o}}$& $45,665$  &$45,697$\\ 
$3\text{s}3\text{p}^3$&${}^3\text{D}_2^{\text{o}}$& $67,231$  &$65,272$\\
$3\text{s}3\text{p}^3 $&${}^3\text{P}_2^{\text{o}}$&  $70,689$ &$76,764$\\
$3\text{s}^23\text{p}3\text{d} $&$ {}^1\text{D}_2^{\text{o}}$&$81,792$ &$77,710$\\
$3\text{s}^23\text{p}3\text{d} $&${}^3\text{D}_2^{\text{o}}$&$91,145$ &$104,102$\\
$3\text{s}^23\text{p}3\text{d} $&$ {}^3\text{F}_2^{\text{o}}$&$96,173$ &$87,804$\\
$3\text{s}3\text{p}^3 $&${}^1\text{D}_2^{\text{o}}$&  $109,604$ &$112,607$\\
$3\text{s}^23\text{p}3\text{d}$&$ {}^3\text{P}_2^{\text{o}}$&$111,287$ &$103,629$\\ \hline\hline \\
\end{tabular}
\end{center}
This table shows that the largest disagreement between fitted and experimental energy levels is approximately $12,957$ cm$^{-1}$ ($3\text{s}^23\text{p}3\text{d} \ {}^3\text{D}_2^{\text{o}}$); however, the largest disagreement within the configuration of interest, $3\text{s}3\text{p}^3$, is a respectable $6,075$ cm$^{-1}$ (${}^3\text{P}_2^{\text{o}}$). Thus, despite some flaws, the $\chi^2$ model succeeded at describing the P II system to a precision commensurate with its sophistication.\\ \section{Failure of the Semi-Emprical Method for J=1 Ground Transitions in P II}
It is worth mentioning that similar application of the semi-empirical method to the J=1 case for $3\text{s}^23\text{p}^2-3\text{s}3\text{p}^3$ transitions was not a success. A 7x7 interaction matrix was generated for \end{multicols}
\begin{figure*}[t]
\begin{flalign*}
\mid 3\text{s}3\text{p}^3 \ {}^1\text{D}_2^{\text{o}} \rangle ' &= 0.975026\mid 3\text{s}3\text{p}^3 \ {}^1\text{D}_2^{\text{o}} \rangle - 0.219777\mid 3\text{s}^23\text{p}3\text{d} \ {}^1\text{D}_2^{\text{o}} \rangle  + 0.031831\mid 3\text{s}^23\text{p}3\text{d} \ {}^3\text{P}_2^{\text{o}} \rangle \\ 
\mid 3\text{s}3\text{p}^3 \ {}^3\text{P}_2^{\text{o}} \rangle ' &= 0.991192\mid 3\text{s}3\text{p}^3 \ {}^3\text{P}_2^{\text{o}} \rangle + 0.109902\mid 3\text{s}^23\text{p}3\text{d} \ {}^3\text{P}_2^{\text{o}} \rangle +  0.0694869 \mid 3\text{s}3\text{p}^3 \ {}^3\text{D}_2^{\text{o}} \rangle \\
\mid 3\text{s}3\text{p}^3 \ {}^3\text{D}_2^{\text{o}} \rangle ' &= 0.963482\mid 3\text{s}3\text{p}^3 \ {}^3\text{D}_2^{\text{o}} \rangle - 0.257754 \mid 3\text{s}^23\text{p}3\text{d} \ {}^3\text{D}_2^{\text{o}} \rangle \\
\mid 3\text{s}3\text{p}^3 \ {}^5\text{S}_2^{\text{o}} \rangle ' &= - 0.999931\mid 3\text{s}3\text{p}^3 \ {}^5\text{S}_2^{\text{o}} \rangle
 \end{flalign*}
\caption{Significant components of eigenvectors for $3\text{s}3\text{p}^3$ J=2 levels\setcounter{footnote}{\value{footnote}}\protect\footnotemark
}
\end{figure*}
\footnotetext{All eigenvector components were used in the computation of branching fractions.}
\begin{multicols}{2}
\noindent levels in the $3\text{s}3\text{p}^3$ and $3\text{s}^23\text{p}3\text{d}$ upper states, but it was impossible to find a sufficiently low point in the $\chi^2$ surface. As a result, an appropriate orthonormal eigenvector basis that would diagonalize the matrix to measured energies could not be obtained. At the moment, it is not clear whether this was due to mathematical failure of the model or due to the presence of a significant interaction in the system that was left out of the calculation. The failure of the J=1 model is unfortunate, and it is important to qualify the Slater parameters of the apparently successful J=2 model by noting that it is impossible to compare them against J=1 values thus far. \section{Computation of Semi-Empirical Branching Fractions for J=2 Transitions}As mentioned above, eigenvectors were obtained for the eight levels of the J=2 case after the $\chi^2$ fit produced Slater parameters. The eigenvectors described each IC/CI mixed state as a linear combination of LS basis states. Although the eigenvectors were found to have nonzero components along every element of the LS basis, each eigenvector had no more than four significant components. The most significant components of each $3\text{s}3\text{p}^3$ eigenvector are displayed in Figure 5. The fit determined the $3\text{s}3\text{p}^3 \ {}^5\text{S}_2^{\text{o}}$ level to be essentially pure, so this level was ignored in computation of branching fractions.

In order to compute branching fractions, it was also necessary to characterize the ground state appropriately, which was accomplished using a method formulated by Dr. Curtis \cite{curtis2003}. In this method, the eigenvectors of the $3\text{s}^23\text{p}^2$ configuration were written in terms of semi-empirical singlet-triplet mixing angles: 
\begin{flalign*}
& \mid {}^3\text{P}_1 \rangle' =  \ \mid {}^3 \text{P}_1 \rangle \\
& \mid {}^3\text{P}_2 \rangle' = \cos \theta_2 \mid {}^3\text{P}_2 \rangle - \sin \theta_2 \mid {}^1\text{D}_2 
\rangle \\
& \mid {}^1\text{D}_2 \rangle' = \sin \theta_2 \mid {}^3\text{P}_2 \rangle + \cos \theta_2 \mid {}^1\text{D}_2 \rangle
\end{flalign*} 
In this case, $\theta_2$ was found to be $1.630^{\circ}$ \cite{curtis2000}, and the J=0 levels of the lower configuration were ignored because of the following E1 transition selection rules: \cite{curtis2003}
\begin{flalign*}
&\Delta J= 0, \pm1 \ \text{(no 0 to 0)} \\
&\Delta L= 0, \pm1 \ \text{(no 0 to 0)} \\ & \Delta S=0
\end{flalign*}
The various line strengths from each upper state to each lower state were then computed (see Figure 8). Each line strength $S_\text{ul}$ was computed as the square of the reduced matrix element $D_\text{ul}$: 
\begin{flalign*}
D_\text{ul} = \langle \psi_{J'}^\text{e} \mid \mathcal{D} \mid \psi_{J}^{\text{o}} \rangle =  \sum_{L',\ L,\ S} C_{L'S}A_{LS}\mathcal{M}_{1} \ \ + \\ 
 \sum_{L',\ L,\ S} C_{L'S}B_{LS}\mathcal{M}_{2} 
\end{flalign*}
where $\mathcal{D}$ is the dipole transition operator, $C_{L'S}$ refers to an accessible component\footnote{That is, accessible by E1 transition rules to the upper state component in question.} of the primed lower state bra, $A_{LS}$ refers to a $3\text{s}3\text{p}^3$ component of the unprimed upper state ket $\mid \! \!  \psi_{J}^{\text{o}} \rangle$, and $B_{LS}$ refers to a $3\text{s}^23\text{p}3\text{d}$ component of the upper state ket. The $\mathcal{M}$ factors are the LS-coupling transition elements, displayed in Figure 7.

Each line strength was then corrected for wavelength, using:
\[
\displaystyle
A_{\text{ul}}' \ \ = \ \  \begin{array}{c} S_{\text{ul}} \\ \midrule \lambda_{\text{ul}}^3 \end{array} \ \ = \ \ \begin{array}{c}\langle \psi_{J'}^\text{e} \mid \mathcal{D} \mid \psi_{J}^{\text{o}}\rangle^2 \\ \midrule \lambda_{\text{ul}}^3 \end{array}
\] 
where the prime on $A_{\text{ul}}'$ indicates that this quantity is proportional, but not equal to, the transition rate $A_{\text{ul}}$. Given the corrected line strengths, it was straightforward to compute the branching fractions for each set of transitions: \\
\[
\text{BF}_{\text{ul}} = \frac{A_{\text{ul} }'}{\sum_{i}A_{\text{u}i}'}
\] \\
where the denominator on the right hand side is the sum of line strengths over all lower states that are E1-accessible to the upper state $u$.

Previously-measured experimental values for the branching fractions in question were unavailable; however, it was possible to compare our semi-empirical branching fractions with the theoretical computations of S. S. Tayal and A. Hibbert. The $^3\text{P}_2^{\text{o}}$ branching fractions were also measured at the THIA laboratory (see Figure 6); unfortunately, it was impossible to measure the $^3\text{D}_2^{\text{o}}$ branching fractions using THIA due to an especially large $\tau$, which is largely caused by configuration-interaction transition rate canceling between the $\text{sp}^3$ and spd components of that level. The various branching fractions and their comparisons are presented in the table below. Note: Tayal and Hibbert quoted ${f}_\text{length}$ values; these branching fractions were computed from those.
\end{multicols}
\begin{table}[h]
\begin{center}
\begin{tabular}{ccccccc}
\hline\hline
Transition&J&Tayal BF \cite{tayal2002}&Hibbert BF \cite{hibbert1987}&LS BF&Semi-Emprical BF&Experimental BF\\ [0.0ex]
\hline \\[-2.0 ex]
${}^3\text{P}_J - {}^3\text{D}_2^{\text{o}}$&1&0.790&0.808&0.753&0.691&$-$ \\
${}^3\text{P}_J - {}^3\text{D}_2^{\text{o}}$&2&0.210&0.192&0.247&0.309&$-$ \\
${}^3\text{P}_J - {}^3\text{P}_2^{\text{o}}$&1&0.241&0.242&0.252&0.318&$0.233 \pm 0.0223$ \\
${}^3\text{P}_J - {}^3\text{P}_2^{\text{o}}$&2&0.759&0.758&0.748&0.682&$0.767 \pm 0.0569$ \\
\hline\hline
\end{tabular}
\end{center}
\end{table}
\begin{figure}[h]
\setlength\fboxsep{0pt}
\setlength\fboxrule{0.5pt}
\fbox{\includegraphics[width=140mm]{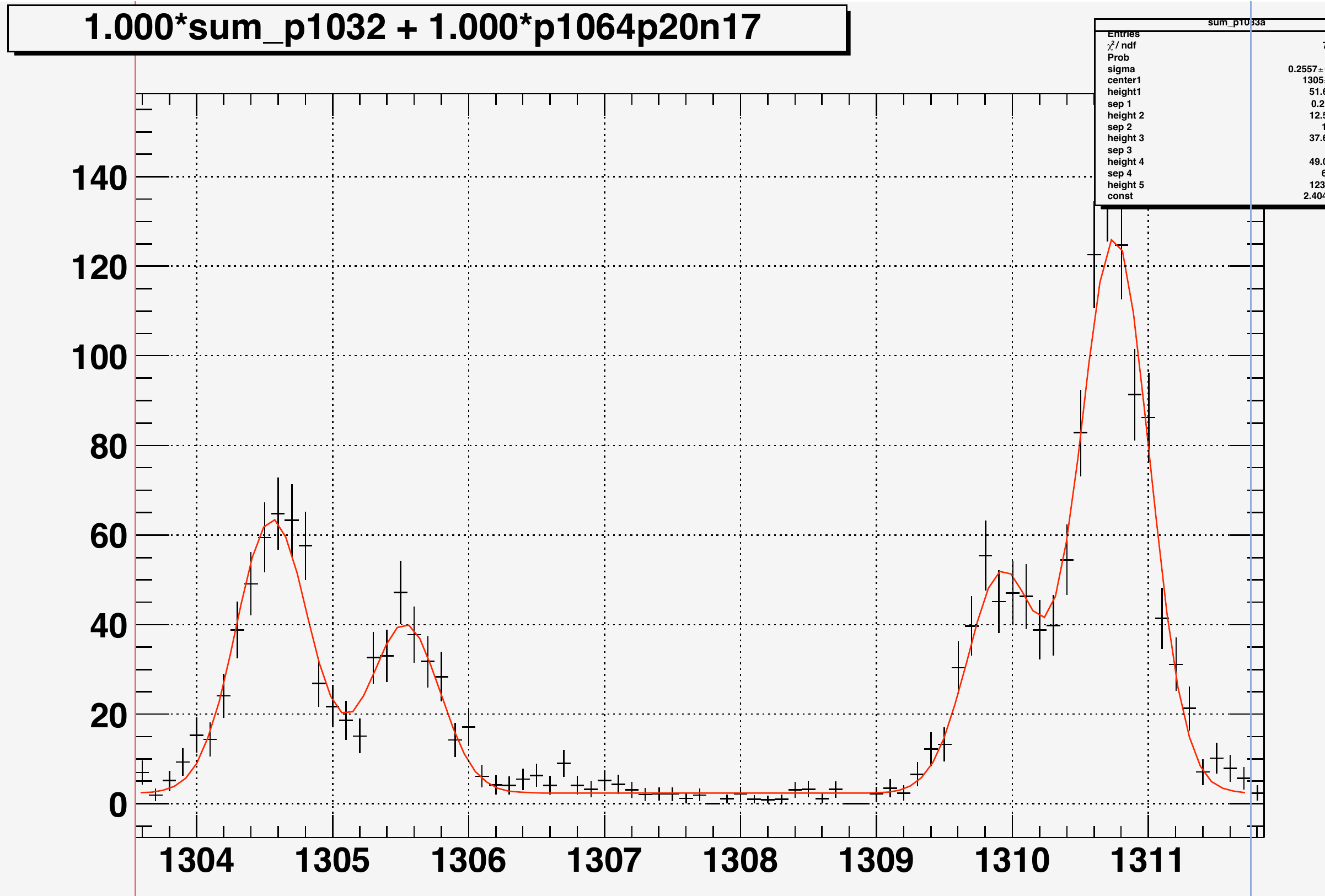}}
\caption{Measurement of  $\text{3s3p}^3 \ {}^3\text{P}_2^{\text{o}}$ branching fractions at THIA. The intensities of the $3\text{s}^2 3\text{p}^2 \ {}^3\text{P}_1\text{ and } ^3\text{P}_2$  lines at 1305.48 $\text{\AA}$ and 1310.70 $\text{\AA}$ were measured using a Gaussian fit upon the sum of multiple spectroscopic scans.}
\end{figure}
\newpage
\begin{figure*}[t]
\begin{small}
\begin{flalign*}
\mathcal{M}_1 =  (-1)^{L'+J+S_2-^1/_2}\sqrt{2}(2L+1)\sqrt{(2J'+1)(2J+1)(2L'+1)(2S_2+1)} \\ \times \left \{ \begin{array}{ccc}L'&S&J'\\J&1&L\end{array}\right \} \left \{ \begin{array}{ccc}S&^1/_2&S_2\\^1/_2&S&0\end{array}\right \} \left \{ \begin{array}{ccc}L'&L&1\\L&L'&0\end{array}\right \} (cfp) \langle 3\text{s} \mid r \mid 3\text{p} \rangle \\ \\
\mathcal{M}_2 =  (-1)^{S+J+1}\sqrt{2(2L'+1)(2L+1)(2J'+1)(2J+1)} \\ \times \left \{ \begin{array}{ccc}L'&S&J'\\J&1&L\end{array}\right \}\left \{ \begin{array}{ccc}L'&L&1\\2&1&1\end{array}\right \} \langle 3\text{p} \mid r \mid 3\text{d} \rangle \\ \\
\langle 3\text{s} \mid r \mid 3\text{p} \rangle = -\mathcal{R}_\text{sp} = -\int_0^\infty P_{3\text{s}}(r)rP_\text{3p}(r)dr \\ \\
\langle 3\text{s} \mid r \mid 3\text{p} \rangle = -\sqrt{2}\mathcal{R}_\text{pd} = -\sqrt{2} \int_0^\infty P_{3\text{p}}(r)rP_\text{3d}(r)dr \\ \\
\begin{array}{llllc}
L'&L&\text{p}^3&S_2&cfp \\
\midrule
1&1&^2\text{P}&^1/_2&-\sqrt{^1/_2} \\
\addlinespace[1pt]
1&2&^2\text{D}&^1/_2&+\sqrt{^1/_2} \\
\addlinespace[1pt]
1&0&^4\text{S}&^3/_2&1 \\
\addlinespace[1pt]
2&1&^2\text{P}&^1/_2&-\sqrt{^5/_{18}} \\
\addlinespace[1pt]
2&2&^2\text{D}&^1/_2&-\sqrt{^1/_2} \\
\addlinespace[1pt]
0&1&^2\text{P}&^1/_2&+\sqrt{^2/_9} \\
\end{array}
\end{flalign*}
\end{small}
\caption{The LS-coupling E1 transition elements, derived from Cowan by Dr. Ellis \cite{cowan1981}. Note that $S_2$ and the coefficient of fractional parentage $cfp$ are given in the table at bottom. For computation of branching fractions, the hydrogenic ratio $\mathcal{R}_\text{pd} = \sqrt{^5/_8}\mathcal{R}_\text{sp} $ was used, so that the radial integrals would cancel.
} \end{figure*}
\begin{figure*}[h]
\begin{small}
\begin{verbatim}
For[i = 1, i < 9, i++,
 For[j = 1, j < 4, j++,
  For[k = 1, k < 9, k++,
   For[l = 1, l < 4, l++,
    If[1 == KroneckerDelta[lowerSLJBasis[[l, 1]], upperSLJBasis[[k, 1]]] 
    && (upperSLJBasis[[k, 2]] == lowerSLJBasis[[l, 2]] || 
        upperSLJBasis[[k, 2]] - 1 == lowerSLJBasis[[l, 2]] || 
        upperSLJBasis[[k, 2]] + 1 == lowerSLJBasis[[l, 2]]) , 
     relativeStrengths[[i, j]] = relativeStrengths[[i, j]] + 
       upperVecs[[i, k]]*lowerVecs[[j, l]]*mSelector[k, upperSLJBasis[[k, 2]], 
       upperSLJBasis[[k, 1]], upperSLJBasis[[k, 3]], lowerSLJBasis[[l, 2]], 
         lowerSLJBasis[[l, 3]]]]     
]]]]
\end{verbatim}
\end{small}
\caption{Computation of all twenty-four  $3\text{s}^2 3\text{p}^2 - \text{3s3p}^3$ J=2 line strengths in Mathematica\setcounter{footnote}{\value{footnote}}\protect\footnotemark}
\end{figure*}
\footnotetext{The outermost loop cycles through all eight upper eigenvectors, the next loop cycles through all three lower eigenvectors, and the inner loops cycle through the components of each upper-lower pair of eigenvectors. The "if" statement imposes selection rules and computes $D_\text{ul}$. }
\begin{multicols}{2}
\section{Conclusion}
The theoretical, semi-empirical, and experimental values presented here are within $0.07$ of pure LS-coupling branching fractions for these levels. This is to be expected, given the high (greater than 92\%) leading components of all upper state eigenvectors in the $\chi^2$ fit. This provides support for Dr. Curtis' hypothesis that little upper state IC, but significant upper state CI, does not cause branching fractions to deviate heavily from LS coupling.

For the branching fractions of the $^3\text{P}_2^{\text{o}}$ upper state, there is good agreement between the THIA values, Tayal, and Hibbert. Similarly, there is good agreement between Tayal and Hibbert for the $^3\text{D}_2^{\text{o}}$ branching fractions. However, the semi-emprical branching fractions quoted here do not agree nearly as well with the other calculations. Due to the good agreement between Tayal, Hibbert, and experiment, it seems likely that the semi-empirical approach is at fault here. Although the semi-empirical method produces branching fractions somewhat close to LS-coupling values, it appears to lack the sophistication necessary to reproduce experiment, or the more advanced theoretical calculations for this system.

The most interesting result of the semi-empirical method as applied to this set of transitions in P II is the complete failure of the J=1 model. Although this outcome could be due to limitations in the model, it could also indicate the presence of a heretofore unnoticed configuration interaction or level crossing present in the system. Further study will be necessary to explore these possibilities.
\end{multicols}
\vspace{2.5cm}
\begin{figure}[h]
\bibliographystyle{plain}
\bibliography{PII}
\end{figure}
\end{document}